\documentclass[conference]{IEEEtran}
\usepackage{hyperref}
\usepackage[T1]{fontenc}
\usepackage[latin9]{inputenc}
\usepackage{amsmath}
\usepackage{amssymb}
\usepackage{graphicx}
\usepackage{esint}

\makeatletter
\usepackage{cite}
\usepackage{psfrag}
\usepackage{url}

\usepackage{array}
\usepackage{epsfig}
\usepackage{multicol}
\usepackage{times}
\IEEEoverridecommandlockouts

\makeatother

\begin{document}

\title{{On Optimal Spectrum Access of Cognitive Relay With Finite Packet
Buffer\thanks{Copyright (c) 2015 IEEE. Personal use of this material is permitted. However, permission to use this material for any other purposes must be obtained from the IEEE by sending a request to pubs-permissions@ieee.org.}\vspace{-0.2cm}
 }}

\author{Kedar Kulkarni,~\IEEEmembership{Student~Member,~IEEE,} and Adrish
Banerjee,~\IEEEmembership{Senior Member,~IEEE}\thanks{The authors are with Department of Electrical Engineering, Indian
Institute of Technology Kanpur, Kanpur 208016, INDIA (e-mail: kulkarni@iitk.ac.in;
adrish@iitk.ac.in).}\vspace{-0.6cm}
 }
\maketitle
\begin{abstract}
We investigate a cognitive radio system where secondary user (SU)
relays primary user (PU) packets using two-phase relaying. SU transmits
its own packets with some access probability in relaying phase using
time sharing. PU and SU have queues of finite capacity which results
in packet loss when the queues are full. Utilizing knowledge of relay
queue state, SU aims to maximize its packet throughput while keeping
packet loss probability of PU below a threshold. By exploiting structure
of the problem, we formulate it as a linear program and find optimal
access policy of SU. We also propose low complexity sub-optimal access
policies, namely constant probability transmission and step transmission.
Numerical results are presented to compare performance of proposed
methods and study effect of queue sizes on packet throughput.\end{abstract}


%\vspace{-0.2cm}

\section{Introduction\label{sec:Introduction}}

In cognitive radio (CR) networks, secondary users (SUs) access spectrum
allocated to primary users (PUs) in such a way that given quality-of-service
(QoS) requirement of PUs is satisfied. Users store packets arriving
from higher layers in queues before transmission over wireless link.
Various works have studied SU packet throughput for non-cooperation
scenarios where SU's access probability is optimized under queue stability
constraint of PU \cite{Simeone_stable_throughput,Jeon,fanous}. Cooperation
between SU and PU improves throughput of both users as shown in \cite{Kompella}
and \cite{Ashour}. These works consider queues of infinite storing
capacity. In practice, queues are of finite size. If a queue is full,
new packets cannot be admitted to the queue and are lost. Queueing
performance of finite sized SU queue was studied in \cite{Chu}. In
\cite{Krikidis_buffer,Schober_buffer2,Schober_buffer,Teh_buffer,Krikidis_buffer_survey},
authors studied relay selection problem for finite buffer-aided relaying
systems. These works considered dedicated relay nodes that do not
have their own data to transmit. In \cite{Shafie} and \cite{Elmahdy},
authors considered cooperative CR networks with finite sized relay
queue and proposed packet admission control assuming that relay queue
information is available at SU. However, underlying assumption in
these works is that PU queue has infinite buffer length. Also, whole
slot is used by the relay either for transmission or for reception.
A relaying protocol where packet reception and transmission takes
place in the same slot using time sharing, may enable the relay to
improve its throughput by transmitting own packets more frequently.

In this paper, we investigate SU throughput in a cooperative CR system
where SU relays failed packets of PU using two-phase relaying. SU
transmits its own packets in the relaying phase using time sharing,
with some access probability. Furthermore, we consider that both PU
and SU have finite capacity queues. SU's finite queue size affects
cooperation offered to PU. Thus, queue sizes at both PU and SU impact
PU's packet loss. Our aim is to find optimal access policy of SU that
maximizes SU packet throughput while satisfying PU's packet loss constraint.
Specifically, our contribution is as follows. 
\begin{itemize}
\item We model PU and relay queues as discrete time Markov chains (DTMC).
Using DTMC analysis, we characterize packet loss probability of PU
and SU packet throughput. 
\item We formulate the problem of maximizing SU throughput under PU packet
loss constraint, which is non-convex. By exploiting structure of the
problem, we transform it into a linear programming (LP) problem over
the feasible range of PU packet throughput. We also propose two low-complexity
suboptimal access methods that transform original multi-dimensional
problem into one dimensional problem. 
\item Finally, we present numerical results to study effect of queue sizes,
path loss and time sharing on SU packet throughput. We also compare
the performance with infinite capacity queue system under queue stability
constraint. 
\end{itemize}

\section{System Model\label{sec:System-Model}}

\begin{figure}
\centering

\includegraphics[scale=0.15]{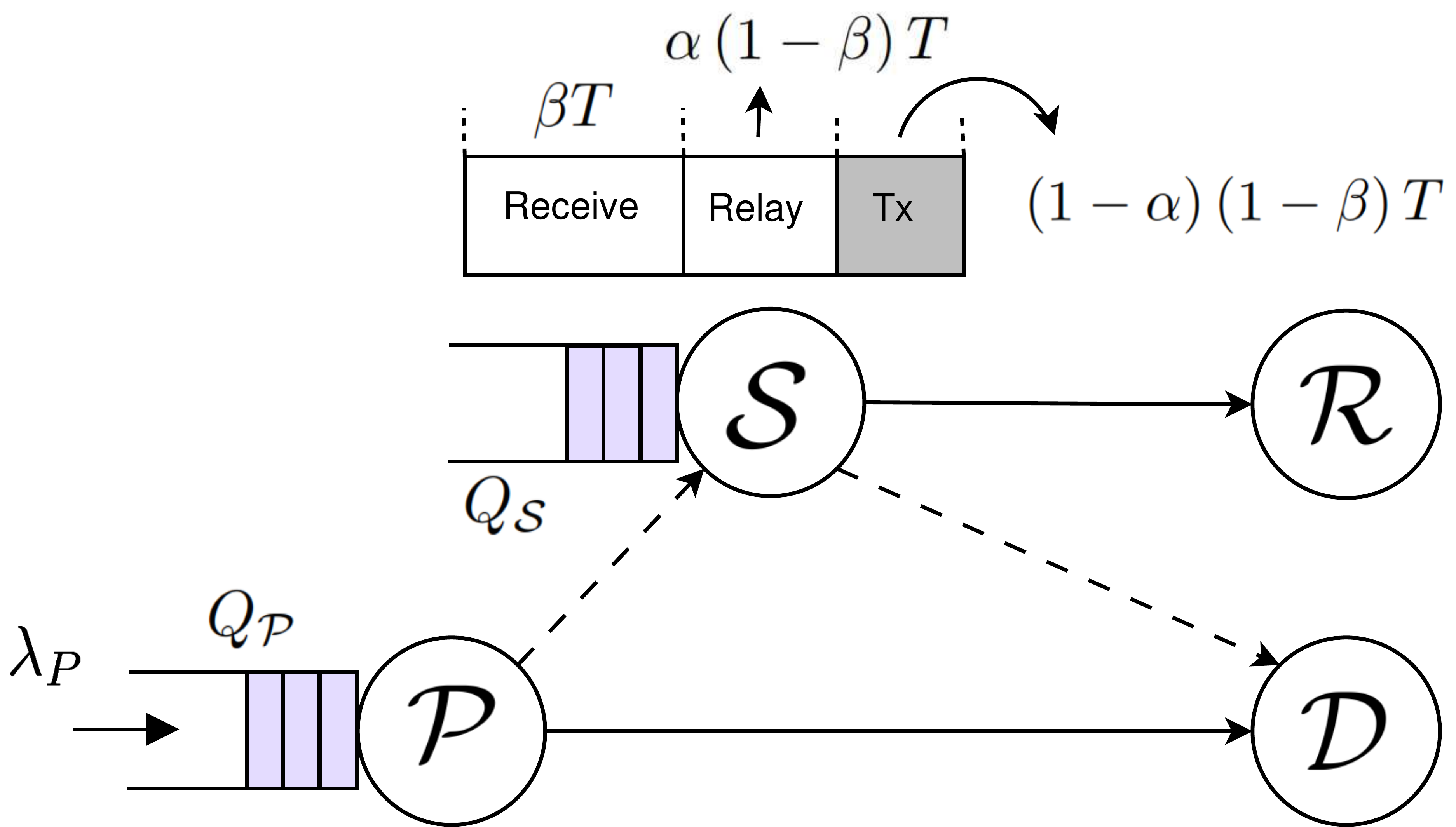}

\protect\protect\protect\protect\protect\protect\caption{System model of SU relaying PU packets and transmitting own packets
using two-phase relaying and time sharing\label{fig:SysMod}}
\vspace{-0.3cm}
\end{figure}

As shown in Fig. \ref{fig:SysMod}, a PU source $\mathcal{P}$ transmits
packets to PU destination $\mathcal{D}$ with assistance of a SU node
$\mathcal{S}$ using two-phase relaying as done in \cite{Schober_buffer2}.
Nodes $\mathcal{P}$ and $\mathcal{S}$ are equipped with packet queue
$Q_{\mathcal{P}}$ of capacity $N_{P}$ and relay queue $Q_{\mathcal{S}}$
of capacity $N_{S}$ respectively. In a slot of duration $T$, $\mathcal{P}$
transmits its packet with power $P_{\mathcal{P}}$ for time $\beta T,\,\beta\in\left[0,\,1\right]$.
If $\mathcal{D}$ fails to receive the packet, it is admitted to the
relay queue at $\mathcal{S}$, provided that the packet is correctly
received at $\mathcal{S}$ and the relay queue is not full. In relaying
phase of duration $\left(1-\beta\right)T$, $\mathcal{S}$ relays
the PU packet to $\mathcal{D}$ with power $P_{\mathcal{S}}$. With
some access probability, SU also transmits its own packets to SU destination
$\mathcal{R}$ using time sharing, that is, SU relays PU packet for
duration $\alpha\left(1-\beta\right)T$ and transmits its own packet
for duration $\left(1-\alpha\right)\left(1-\beta\right)T$, $\alpha\in\left[0,\,1\right]$.
The access probability is $p_{n}$ when there are $n,\,0\leq n\leq N_{S}$
packets in $Q_{\mathcal{S}}$. If $Q_{\mathcal{S}}$ is empty, whole
relaying duration $\left(1-\beta\right)T$ is used to transmit SU
packet, with probability $p_{0}=1$.

We assume that all channels are independent block-fading in nature,
that is, channel gains remain constant during a slot and vary independently
from slot to slot. Channel power gain between source $s$ and destination
$d$ is denoted as $g_{sd}$ and is exponentially distributed with
mean $\sigma_{sd}^{2},\,s,d\in\left\{ \mathcal{P},\,\mathcal{S},\,\mathcal{D},\,\mathcal{R}\right\} $.
The distance between $s$ and $d$ is denoted by $r_{sd}$ and path-loss
exponent is denoted by $\kappa$. Additive white Gaussian noise (AWGN)
at receivers has power $\sigma_{\mathcal{N}}^{2}$. PU and SU packets
have fixed length of $\mathcal{B}$ bits. A packet is assumed to be
delivered successfully to intended receiver if instantaneous channel
capacity is greater than required transmission rate. Then probability
of successful packet transmission is given by \cite{fanous} \vspace{-0.2cm}
 
\begin{equation}
\theta_{sd}\!=\!\Pr\!\left[\!\log_{2}\left(1+\frac{g_{sd}P_{s}r_{sd}^{-\kappa}}{\sigma_{\mathcal{N}}^{2}}\right)\!\geq\!\frac{\mathcal{B}}{WT_{s}}\!\right]\!\!=\!\exp\!\left(\!\!\frac{\!-\sigma_{\mathcal{N}}^{2}\!\left(\!2^{\frac{\mathcal{B}}{WT_{s}}}\!-\!1\!\right)}{P_{s}r_{sd}^{-\kappa}\sigma_{sd}^{2}}\!\!\right)\!\!,
\end{equation}
where $P_{s}$ is transmit power, $T_{s}$ is transmission duration
and $W$ is channel bandwidth. We denote probability of successful
packet transmission on $s-d$ link without time sharing by $\theta_{sd},\,s,d\in\left\{ \mathcal{P},\,\mathcal{S},\,\mathcal{D},\,\mathcal{R}\right\} $.
In case of time sharing, time available for relaying/transmission
of PU and SU packets is less than $\left(1-\beta\right)T$. We use
$\overline{\theta_{\mathcal{SD}}}$ and $\overline{\theta_{\mathcal{SR}}}$
to denote successful transmission probabilities in case of time sharing\footnote{Note that notation $\overline{\theta_{sd}}$ only signifies success
probability with time sharing and $\overline{\theta_{sd}}\neq1-\theta_{sd}$.}. As required transmission rate is higher, probabilities of successful
packet transmission decrease. Thus, we have $\overline{\theta_{\mathcal{SD}}}<\theta_{\mathcal{SD}}$
and $\overline{\theta_{\mathcal{SR}}}<\theta_{\mathcal{SR}}$. Successful
transmission probabilities on all links are known to the SU \cite{Jeon,fanous,Kompella}.
\vspace{-0.1cm}
 
\subsection{Queue blocking and packet loss}

Packet arrival process at PU queue $Q_{\mathcal{P}}$ is Bernoulli
with average rate $\lambda_{P}\in\left[0,\,1\right]$ packets/slot.
A packet is removed from $Q_{\mathcal{P}}$ only when it is received
at $\mathcal{D}$ or $\mathcal{S}$. A PU packet is admitted to the
relay queue $Q_{\mathcal{S}}$ when all of the following events are
true-- 1) Packet transmission on $\mathcal{P}-\mathcal{D}$ link fails,
2) PU packet is successfully received at $\mathcal{S}$, and 3) $Q_{\mathcal{S}}$
is not full. Thus, packet departure rate at $Q_{\mathcal{P}}$, denoted
as $\mu_{P}$, depends on channel between $\mathcal{P}-\mathcal{S}$
and state of $Q_{\mathcal{S}}$. When $Q_{\mathcal{P}}$ is full,
new packets cannot be admitted to the queue and are dropped.

\begin{figure}
\centering

\includegraphics[width=8.2cm,height=5.4cm]{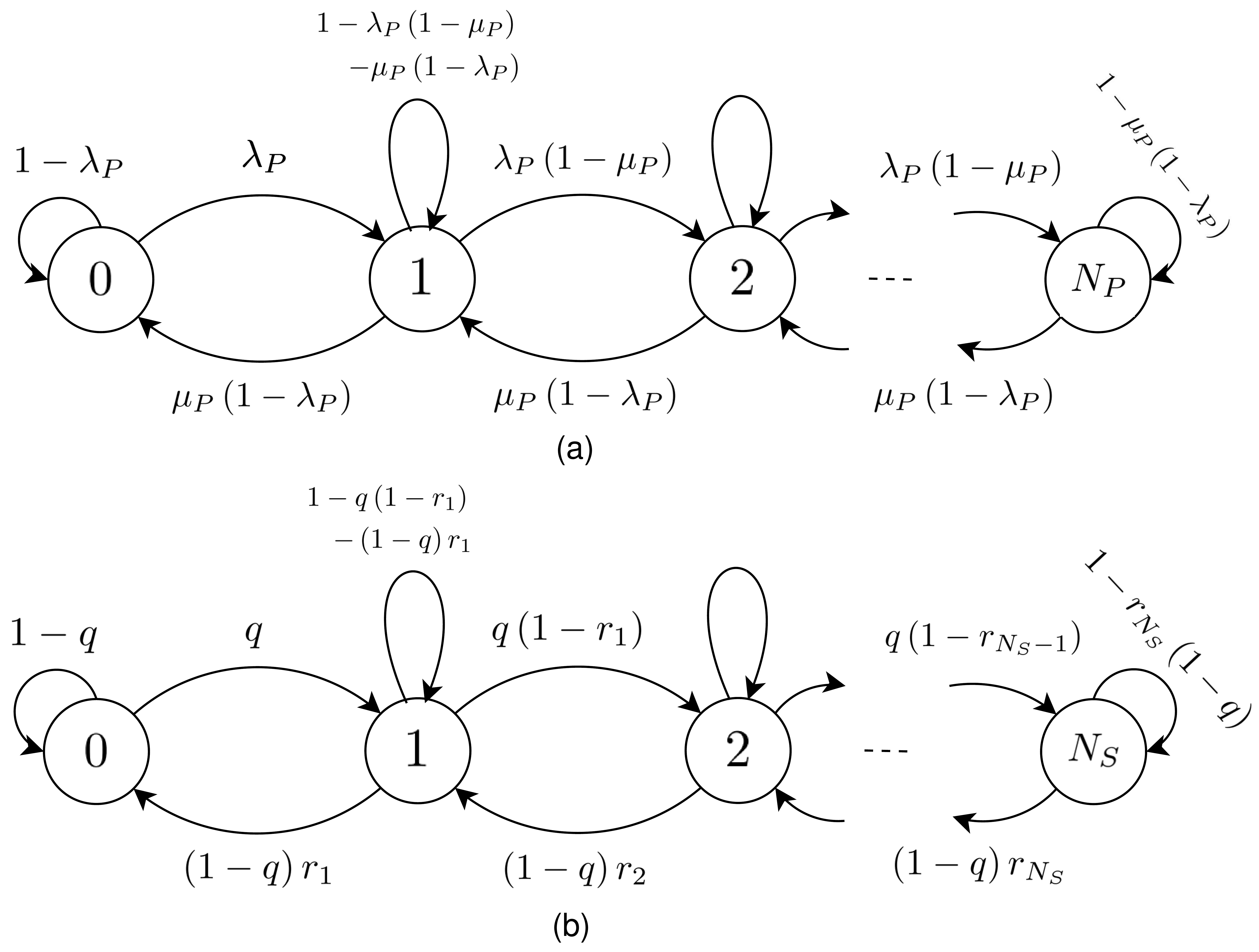}

\caption{Discrete time Markov chain (DTMC) model of (a) PU queue $Q_{\mathcal{P}}$
and (b) relay queue $Q_{\mathcal{S}}$\label{fig:DTMC}}

\vspace{-0.3cm}
\end{figure}

PU queue can be modeled as a discrete time Markov chain (DTMC) as
shown in Fig \ref{fig:DTMC}(a) where states denote number of packets
in PU queue. Let $w_{n},\,n=0,\,1,\dots,\,N_{P}$ be steady state
probability of PU queue being in state $n$. Also let $\gamma=\frac{\lambda_{P}\left(1-\mu_{P}\right)}{\left(1-\lambda_{P}\right)\mu_{P}}$.
Then we can write local balance equations for DTMC of $Q_{\mathcal{P}}$
as 
\begin{align}
w_{1} & =\frac{\gamma}{\left(1-\mu_{P}\right)}w_{0},\\
w_{n+1} & =\gamma\,w_{n},\,n=1,\,2,\dots,\,N_{P}-1.
\end{align}
Noting that $w_{n}=\gamma^{n-1}w_{1},\,n>1$ and $\sum_{n=0}^{N_{P}}w_{n}=1$,
we get probability of $Q_{\mathcal{P}}$ being empty as 
\[
w_{0}=\begin{cases}
\frac{\left(1-\mu_{P}\right)\left(1-\gamma\right)}{1-\mu_{P}\left(1-\gamma\right)-\gamma^{N_{P}+1}} & \,\,\mbox{for}\,\,\gamma\neq1\\
\frac{1-\mu_{P}}{N_{P}+1-\mu_{P}} & \,\,\mbox{for}\,\,\gamma=1.
\end{cases}
\]

Then probabilities of $Q_{\mathcal{P}}$ being non-empty and $Q_{\mathcal{P}}$
being full are given by $\nu_{1}=1-w_{0}$ and $\nu_{N_{P}}=w_{N_{P}}=\frac{1}{\left(1-\mu_{P}\right)}\gamma^{N_{P}}w_{0}$
respectively. To keep PU packet loss below a limit, probability of
$Q_{\mathcal{P}}$ being full should remain below a threshold $\epsilon$,
i.e. $\nu_{N_{P}}\leq\epsilon$. From Fig. \ref{fig:DTMC}(a), we
observe that, any increase in $\mu_{P}$ would decrease probability
of $Q_{\mathcal{P}}$ being full, that is, $\nu_{N_{P}}$ monotonically
decreases in $\mu_{P}$. Thus, for given value of $\lambda_{P}$,
we can find $\overline{\mu_{P}}\in\left[0,\,1\right]$ such that $\nu_{N_{P}}\left(\overline{\mu_{P}}\right)=\epsilon$,
using Bisection method. Packet loss constraint $\nu_{N_{P}}\leq\epsilon$
can now be written as $\mu_{P}\geq\overline{\mu_{P}}$. SU should
choose its access probability in such a way that packet loss constraint
of PU is satisfied.

\section{Optimal spectrum access}

DTMC of the relay queue $Q_{\mathcal{S}}$ is as shown in Fig. \ref{fig:DTMC}(b)
where state $n$ denotes number of packets in relay queue at the end
of receiving phase. Probability of a PU packet arriving at $Q_{\mathcal{S}}$
is $q$. When $Q_{\mathcal{S}}$ is in state $n$, SU transmits its
own packets with probability $p_{n}$ using time sharing. Thus, with
probability $\left(1-p_{n}\right)$, PU packet is relayed for duration
$\left(1-\beta\right)T$ and with probability $p_{n}$, PU packet
is relayed for duration $\alpha\left(1-\beta\right)T$. Then probability
of a PU packet departing $Q_{\mathcal{S}}$ in state $n>0$ is
\begin{equation}
r_{n}=\left(1-p_{n}\right)\theta_{\mathcal{SD}}+p_{n}\overline{\theta_{\mathcal{SD}}}=\theta_{\mathcal{SD}}-p_{n}\left(\theta_{\mathcal{SD}}-\overline{\theta_{\mathcal{SD}}}\right).\label{eq:mu_relay}
\end{equation}

When PU is present, a packet is received at $\mathcal{S}$ with probability
$\left(1-\theta_{\mathcal{PD}}\right)\theta_{\mathcal{PS}}$. Thus,
we have $q=\nu_{1}\theta_{\mathcal{PS}}\left(1-\theta_{\mathcal{PD}}\right)$.
For $1\leq n<N_{S}$, state transition from $n$ to $\left(n+1\right)$
occurs when packet transmission of a packet in $Q_{\mathcal{S}}$
fails and a new PU packet is received, which happens with probability
$q\left(1-r_{n}\right)$. State transition from $n$ to $\left(n-1\right)$
occurs when a packet is successfully relayed and no new packet arrives,
which happens with probability $\left(1-q\right)r_{n}$. Let $\pi_{n},\,n=0,\,1,\dots,\,N_{S}$
be steady state probability of $Q_{\mathcal{S}}$ being in state $n$.
Then we write local balance equations as 
\begin{align}
\pi_{1} & =\frac{q}{\left(1-q\right)r_{1}}\pi_{0},\label{eq:balance1}\\
\pi_{n+1} & =\frac{q\left(1-r_{n}\right)}{\left(1-q\right)r_{n+1}}\pi_{n},\,\,n=1,\,2,\dots,\,N_{S}-1.\label{eq:balance2}
\end{align}
For given values of $\lambda_{P}$ and $\mu_{P}$, steady state probabilities
of relay queue can be calculated from (\ref{eq:balance1}), (\ref{eq:balance2})
using 
\begin{equation}
\sum_{n=0}^{N_{S}}\pi_{n}=1.\label{eq:normalize}
\end{equation}

At the start of receiving phase, $Q_{\mathcal{S}}$ is full with probability
$\pi_{N_{S}}\left(1-r_{N_{S}}\right)$. As a PU packet is admitted
to $Q_{\mathcal{S}}$ only when $Q_{\mathcal{S}}$ is not full, we
obtain packet departure rate of PU queue as 
\begin{equation}
\mu_{P}=\theta_{\mathcal{PD}}+\theta_{\mathcal{PS}}\left(1-\theta_{\mathcal{PD}}\right)\left[1-\pi_{N_{S}}\left(1-r_{N_{S}}\right)\right].\label{eq:mup_relay}
\end{equation}

\subsection{SU throughput maximization}

When there are $n>0$ packets in $Q_{\mathcal{S}}$, SU transmits
its own packet for duration $\left(1-\alpha\right)\left(1-\beta\right)T$
with probability $p_{n}$. If $Q_{\mathcal{S}}$ is empty, whole duration
$\left(1-\beta\right)T$ is used to transmit SU packet with probability
$p_{0}=1$. Given PU packet arrival rate $\lambda_{P}$, our objective
is to maximize SU packet throughput while ensuring that packet loss
probability of PU is kept below specified threshold. Thus, the optimization
problem is written as 
\begin{align}
\max_{\boldsymbol{p},\,\boldsymbol{\pi},\,\mu_{P}} & \quad\mu_{S}=\theta_{\mathcal{SR}}\pi_{0}+\overline{\theta_{\mathcal{SR}}}\sum_{n=1}^{N_{S}}\pi_{n}p_{n}\label{eq:Opt_problem}\\
\mbox{s. t.} & \quad\mu_{P}\geq\overline{\mu_{P}},\label{eq:blocking_constraint}\\
 & \quad0\leq p_{n},\pi_{n}\leq1,\,\,n=0,\,1,\dots,\,N_{S},\nonumber \\
 & \quad p_{0}=1\nonumber \\
 & \quad\mu_{P}=\theta_{\mathcal{PD}}+\theta_{\mathcal{PS}}\left(1-\theta_{\mathcal{PD}}\right)\left[1-\pi_{N_{S}}\left(1-r_{N_{S}}\right)\right],\nonumber \\
 & \quad(\ref{eq:balance1}),\,(\ref{eq:balance2}),\,(\ref{eq:normalize}),\nonumber 
\end{align}
where $\boldsymbol{p}=\left[p_{0},\dots,\,p_{N_{S}}\right]^{T}$ and
$\boldsymbol{\pi}=\left[\pi_{0},\dots,\,\pi_{N_{S}}\right]^{T}$.
Optimization problem in (\ref{eq:Opt_problem}) is non-convex due
to product terms of optimization variables $\pi_{n}$ and $p_{n}$.
We transform it into a linear programming (LP) problem by exploiting
structure of the problem.

Let $a_{n}=\pi_{n}p_{n}$. Then we have $a_{0}=\pi_{0}$ and $0\leq a_{n}\leq\pi_{n},\,\,n=1,\,2,\dots,\,N_{S}$.
From (\ref{eq:normalize}), we have 
\begin{equation}
0\leq\sum_{n=0}^{N_{S}}a_{n}\leq1.\label{eq:normalize_new}
\end{equation}
Using (\ref{eq:mu_relay}), we can transform balance equations (\ref{eq:balance1})
and (\ref{eq:balance2}) as given in (\ref{eq:balance_new_1}) and
(\ref{eq:balance_new_2}) on next page.

\begin{figure*}
\begin{equation}
\theta_{\mathcal{SD}}\left(1-q\right)\pi_{1}-q\pi_{0}=\left(\theta_{\mathcal{SD}}-\overline{\theta_{\mathcal{SD}}}\right)\left(1-q\right)a_{1},\label{eq:balance_new_1}
\end{equation}
\begin{equation}
\theta_{\mathcal{SD}}\left(1-q\right)\pi_{n+1}-q\left(1-\theta_{\mathcal{SD}}\right)\pi_{n}=\left(\theta_{\mathcal{SD}}-\overline{\theta_{\mathcal{SD}}}\right)\left(1-q\right)a_{n+1}+q\left(\theta_{\mathcal{SD}}-\overline{\theta_{\mathcal{SD}}}\right)a_{n},\,\,n=1,\dots,\,N_{S}-1.\label{eq:balance_new_2}
\end{equation}

\vspace{-0.2cm}

\rule[0.5ex]{2\columnwidth}{0.5pt}\vspace{-0.6cm}
\end{figure*}

Similarly, constraint in (\ref{eq:mup_relay}) can be written as 
\begin{equation}
\left(1-\theta_{\mathcal{SD}}\right)\pi_{N_{S}}+\left(\theta_{\mathcal{SD}}-\overline{\theta_{\mathcal{SD}}}\right)a_{N_{S}}=1-\frac{\mu_{P}-\theta_{\mathcal{PD}}}{\theta_{\mathcal{PS}}\left(1-\theta_{\mathcal{PD}}\right)}.\label{eq:mup_relay_new}
\end{equation}
Thus, constraints (\ref{eq:balance1}), (\ref{eq:balance2}), (\ref{eq:mup_relay})
are transformed into constraints (\ref{eq:normalize_new}), (\ref{eq:balance_new_1}),
(\ref{eq:balance_new_2}) and (\ref{eq:mup_relay_new}) which are
affine in $\pi_{n}$ and $a_{n}$. The optimization problem in (\ref{eq:Opt_problem})
is still non-convex in $\mu_{P}$. However, for a given value of $\mu_{P}$,
the problem becomes a LP problem in variables $\boldsymbol{\pi}$
and $\boldsymbol{a}=\left[a_{0},\,a_{1},\dots,\,a_{N_{S}}\right]^{T}$
and is written as 
\begin{align}
\max_{\boldsymbol{\pi},\,\boldsymbol{a}} & \quad\theta_{\mathcal{SR}}a_{0}+\overline{\theta_{\mathcal{SR}}}\sum_{n=1}^{N_{S}}a_{n}\label{eq:opt_prob_new}\\
\mbox{s. t.} & \quad0\leq\pi_{n}\leq1,\,\,n=0,\thinspace1,\dots,\,N_{S},\nonumber \\
 & \quad a_{0}=\pi_{0},\,\,0\leq a_{n}\leq\pi_{n},\,\,n=1,\,2,\dots,\,N_{S},\nonumber \\
 & \quad(\ref{eq:normalize}),\,(\ref{eq:normalize_new}),\,(\ref{eq:balance_new_1}),\,(\ref{eq:balance_new_2}),\,(\ref{eq:mup_relay_new}).\nonumber 
\end{align}
From (\ref{eq:mup_relay}) and (\ref{eq:blocking_constraint}), we
see that the feasible values of $\mu_{P}$ are
\begin{align}
\max\left\{ \overline{\mu_{P}},\,\theta_{\mathcal{PD}}+\theta_{\mathcal{PS}}\overline{\theta_{\mathcal{SD}}}\left(1-\theta_{\mathcal{PD}}\right)\right\} \nonumber \\
 & \hspace{-3cm}\leq\mu_{P}\leq\theta_{\mathcal{PD}}+\theta_{\mathcal{PS}}\left(1-\theta_{\mathcal{PD}}\right).\label{eq:feasible}
\end{align}
 The linear program in (\ref{eq:opt_prob_new}) is solved over feasible
values of $\mu_{P}$. Value of $\mu_{P}$ that corresponds to the
maximum SU packet throughput is chosen. From optimal $a_{n}$ and
$\pi_{n}$, optimal SU access probabilities are found as $p_{n}=\frac{a_{n}}{\pi_{n}},\,\,n=0,\,1,\dots,\,N_{S}$.
We have used CVX package for MATLAB \cite{cvx} to solve (\ref{eq:opt_prob_new})
in polynomial complexity.

\subsection{Suboptimal methods}

From (\ref{eq:mu_relay}), (\ref{eq:balance1}), (\ref{eq:balance2})
and (\ref{eq:normalize}), we get steady state probabilities for $Q_{\mathcal{S}}$
as 
\begin{align}
\pi_{0} & =\left[1+\frac{1}{r_{1}}\sum_{n=1}^{N_{S}}\left(\frac{q}{1-q}\right)^{n}\prod_{m=1}^{n-1}\left(\frac{1-r_{m}}{r_{m+1}}\right)\right]^{-1},\label{eq:relay_empty}\\
\pi_{n} & =\left[\left(\frac{q}{1-q}\right)^{n}\frac{1}{r_{1}}\prod_{m=1}^{n-1}\left(\frac{1-r_{m}}{r_{m+1}}\right)\right]\pi_{0},\,\,n>0.
\end{align}
It can be proven that $\pi_{0}$ is monotonically decreasing function
of access probability $p_{n},\,n=1,\,2,\dots,\,N_{S}$. Also we can
prove that $\pi_{n},\,0<n\leq N_{S}$ is a monotonically increasing
function of $p_{m},\,m\leq n$ and a monotonically decreasing function
of $p_{m},\,m>n$. Intuitively, this can be explained from Fig. \ref{fig:DTMC}(b)
as follows. As access probability $p_{m}$ increases, packet departure
rate of $Q_{\mathcal{S}}$ decreases. Thus, more packets get queued
up in $Q_{\mathcal{S}}$. Hence, the probability of $Q_{\mathcal{S}}$
having more than $m$ packets increases, while probability of $Q_{\mathcal{S}}$
having packets less than or equal to $m$ decreases. Using this nature,
we propose low complexity suboptimal methods that simplify $\left(N_{S}+1\right)$
dimensional problem in (\ref{eq:Opt_problem}) to a one-dimensional
problem.

\subsubsection{Constant probability transmission (CPT)}

In this method, SU transmits its own packets with a fixed probability
$p$ when there are $n>0$ packets in relay queue. Thus, we have 
\begin{equation}
p_{n}=\begin{cases}
1 & \,\mbox{for}\,n=0\\
p & \,\mbox{otherwise}.
\end{cases}\label{eq:prob_const_trans}
\end{equation}
In this case, SU packet throughput is $\mu_{S}=\theta_{\mathcal{SR}}\pi_{0}+\overline{\theta_{\mathcal{SR}}}p\sum_{n=1}^{N_{S}}\pi_{n}$.
Using $\sum_{n=1}^{N_{S}}\pi_{n}=1-\pi_{0}$, we can write the problem
of maximizing $\mu_{S}$ for fixed $\mu_{P}$ as 
\begin{align*}
\max_{p\in\left[0,\,1\right]} & \quad\overline{\theta_{\mathcal{SR}}}p+\pi_{0}\left(\theta_{\mathcal{SR}}-\overline{\theta_{\mathcal{SR}}}p\right)\\
\mbox{s. t.} & \quad(\ref{eq:mup_relay}),\,(\ref{eq:balance1}),\,(\ref{eq:balance2}),\,(\ref{eq:normalize}),\,(\ref{eq:prob_const_trans}).
\end{align*}
The term $\pi_{0}\left(\theta_{\mathcal{SR}}-\overline{\theta_{\mathcal{SR}}}p\right)$
is monotonically decreasing in $p$ while term $\overline{\theta_{\mathcal{SR}}}p$
is increasing in $p$. Thus, there exists a unique $p$ that maximizes
$\mu_{S}$. Optimal solution can be found using Interval halving method
with complexity $\mathcal{O}\left(1\right)$.

\subsubsection{Step transmission (ST)}

In this method, SU transmits its own packets using time sharing with
probability $1$ until length of $Q_{\mathcal{S}}$ reaches a threshold
$N_{th}$. Once it crosses $N_{th}$, the relaying phase duration
of $\left(1-\beta\right)T$ is used only to relay PU packets. Thus,
we have 
\begin{equation}
p_{n}=\begin{cases}
1 & \,\mbox{for}\,n\leq N_{th}\\
0 & \,\mbox{otherwise}.
\end{cases}\label{eq:prob_step_trans}
\end{equation}
In this case, the objective is 
\begin{align*}
\max_{N_{th}\in\left\{ 0,\,1,\dots,\,N_{S}\right\} } & \quad\theta_{\mathcal{SR}}\pi_{0}+\overline{\theta_{\mathcal{SR}}}\sum_{n=1}^{N_{th}}\pi_{n}\\
\mbox{s. t.} & \quad(\ref{eq:balance1}),\,(\ref{eq:balance2}),\,(\ref{eq:normalize}),\,(\ref{eq:mup_relay}),\,(\ref{eq:prob_step_trans}).
\end{align*}
With increasing $N_{th}$, $\pi_{0}$ decreases while number of terms
in summation increase. If value of $\overline{\theta_{\mathcal{SR}}}$
is very low compared to $\theta_{\mathcal{SR}}$, decrease in $\pi_{0}$
is significant and $\mu_{S}$ initially decreases. But as $\pi_{0}$
approaches zero, $\mu_{S}$ increases due to increasing value of $\overline{\theta_{\mathcal{SR}}}\sum_{n=1}^{N_{th}}\pi_{n}$.
For high value of $\overline{\theta_{\mathcal{SR}}}$, $\mu_{S}$
increases with increasing $N_{th}$. Throughput drops to zero when
$\pi_{N_{S}}$ increases to such a value that constraint (\ref{eq:mu_relay})
cannot be satisfied. Value of $N_{th}$ that maximizes $\mu_{S}$
can be found by linear search with complexity $\mathcal{O}\left(N_{S}\right)$.

Suboptimal methods run over all feasible values of $\mu_{P}$ given
in (\ref{eq:feasible}) and the value that corresponds to maximum
SU packet throughput is chosen.

\section{Numerical results and discussion}

Parameter values used to plot results are as follows. Transmit power
is $P_{\mathcal{P}}=P_{\mathcal{S}}=0.1\,\mbox{W}.$ Frame duration
is $T=100\,\mbox{ms}$. Time sharing factors are $\beta=\alpha=0.5$
unless stated otherwise. All channels have average channel gain $\sigma_{sd}^{2}=-10\,\mbox{dB},\,s,d\in\left\{ \mathcal{P},\,\mathcal{S},\,\mathcal{D},\,\mathcal{R}\right\} $.
Noise power is $\sigma_{\mathcal{N}}^{2}=10^{-5}\,\mbox{W}$. We take
$\mathcal{B}/W=3\times10^{-3}\,\mbox{bits/Hz}$. We consider $r_{\mathcal{PS}}=r_{\mathcal{SD}}=r_{\mathcal{SR}}=100\,\mbox{m}$.
Path loss exponent is $\kappa=2$. Packet loss probability threshold
is $\epsilon=0.01$.

\subsubsection{Throughput region}

\begin{figure}
\centering

\includegraphics[scale=0.44]{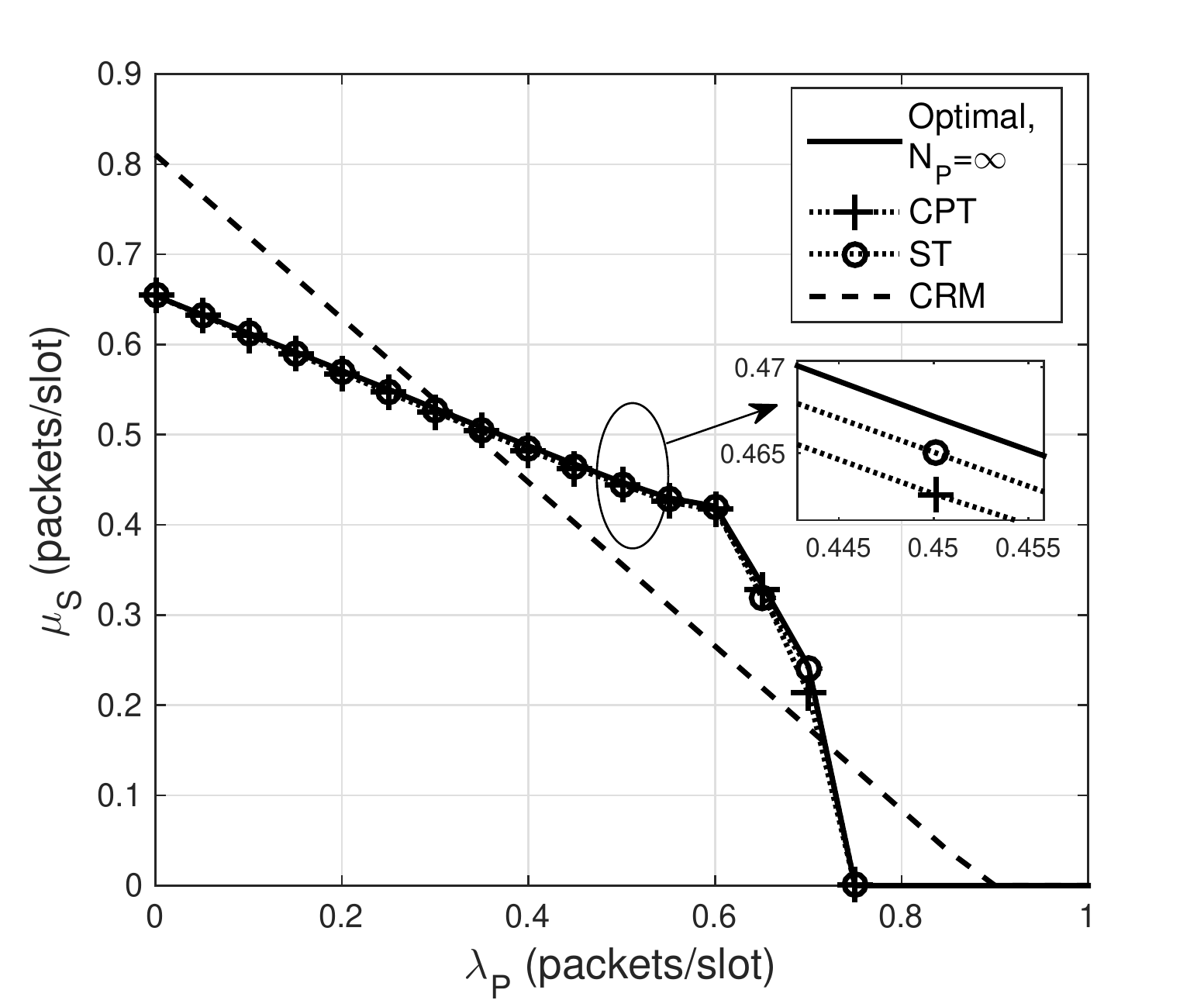} \vspace{-0.2cm}
 \protect\protect\protect\caption{SU packet throughput $\mu_{S}$ versus PU packet arrival rate $\lambda_{P}$
for $N_{S}=10$ and $N_{P}\rightarrow\infty$. \label{fig:SUvsLem_eps}}

\vspace{-0.4cm}
 
\end{figure}

Fig. \ref{fig:SUvsLem_eps} plots throughput region of proposed cooperation
model. As $\lambda_{P}$ increases, higher $\mu_{P}$ is required
to satisfy PU packet loss constraint. To support high $\mu_{P}$,
SU lowers its access probability. Thus, $\mu_{S}$ decreases with
increase in $\lambda_{P}$. As $\lambda_{P}$ increases further, constraint
(\ref{eq:blocking_constraint}) becomes infeasible, at which point
$\mu_{S}$ drops to zero. We also see that performance of constant
probability transmission (CPT) method and step transmission (ST) method
is close to that of optimal method. Hence, the suboptimal methods
are good low-complexity alternatives to the optimal method. 

As a baseline for comparison, we also plot throughput region of cooperative
relaying method (CRM) in \cite{Elmahdy}. In CRM, PU utilizes whole
undivided frame duration $T$ for transmission/reception and optimizes
SU access probability under PU queue stability constraint. In contrast,
two-phase relaying model dedicates $\beta T$ duration for reception
in each slot. Thus, in CRM, probabilities of successful transmission
on $\mathcal{P}-\mathcal{D}$ and $\mathcal{S}-\mathcal{R}$ links
are higher, resulting in better performance of CRM at low and high
values of $\lambda_{P}$. But in mid-range of $\lambda_{P}$, two-phase
relaying benefits by gaining time to transmit own packets as SU relays
PU packets in the same slot.

\subsubsection{Effect of queue sizes}

\begin{figure}
\centering

\includegraphics[scale=0.50]{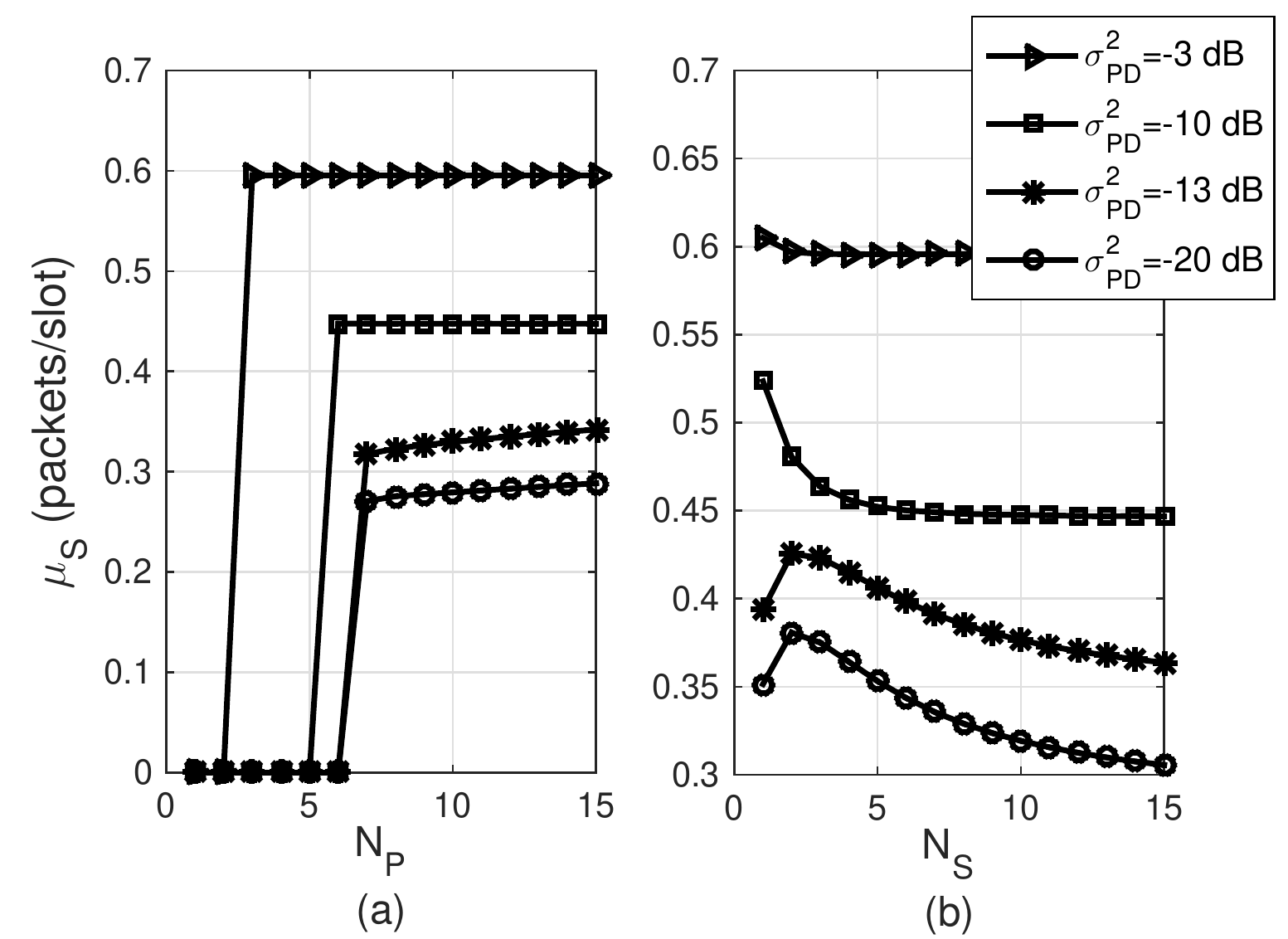} \vspace{-0.2cm}
 \protect\protect\protect\caption{Effect of queue sizes on SU packet throughput $\mu_{S}$ for (a) $N_{S}=10$
(b) $N_{P}=100$ and $\lambda_{P}=0.5$ \label{fig:SUvsLem_NP}}

\vspace{-0.4cm}
 
\end{figure}

Fig. \ref{fig:SUvsLem_NP}(a) plots SU packet throughput $\mu_{S}$
against PU queue capacity $N_{P}$ for different values of $\mathcal{P}-\mathcal{D}$
channel gains. Low values of $N_{P}$ cannot satisfy packet loss constraint
in (\ref{eq:blocking_constraint}). Packet throughput achieved in
such infeasible cases in zero. Increasing $N_{P}$ decreases $\overline{\mu_{P}}$
which is the minimum PU departure rate required to satisfy packet
loss constraint. This allows SU to transmit its own packets with higher
access probabilities. Thus, $\mu_{S}$ increases with increase in
$N_{P}$. As $N_{P}$ increases further, decrease in $\overline{\mu_{P}}$
is insignificant. Access probabilities of SU become constant and in
turn $\mu_{S}$ becomes constant. For high value of $\sigma_{\mathcal{PD}}^{2}$,
PU packet arrival rate at $Q_{\mathcal{S}}$ is less which allows
higher SU access probabilities. Thus, $\mu_{S}$ increases as $\sigma_{\mathcal{PD}}^{2}$
increases.

Fig. \ref{fig:SUvsLem_NP}(b) shows an interesting tradeoff involving
relay queue capacity $N_{S}$. Increase in $N_{S}$ allows SU to transmit
its own packets with higher probability. Also, from (\ref{eq:relay_empty}),
we observe that increase in $N_{S}$ decreases probability of relay
queue being empty $\pi_{0}$. For high values of $\sigma_{\mathcal{PD}}^{2}$,
PU packet arrival rate at $Q_{\mathcal{S}}$ is less. In this case,
decrease in $\pi_{0}$ (and subsequent decrease in $\theta_{\mathcal{SR}}\pi_{0}$)
is significant compared to increase in $\mu_{S}$ due to higher access
probability. Thus, $\mu_{S}$ decreases with increase in $N_{S}$.
For low values of $\sigma_{\mathcal{PD}}^{2}$, PU packet arrival
rate at $Q_{\mathcal{S}}$ is more. In this case, increase in SU throughput
due to higher access probability is significant. But as $N_{S}$ increases
further, $\pi_{0}$ approaches zero and $\sum_{n=1}^{N_{S}}\pi_{n}p_{n}$
decreases. Thus, with increasing $N_{S}$, $\mu_{S}$ initially increases
and then gradually decreases.

\subsubsection{Effect of distance}

\begin{figure}
\centering

\includegraphics[scale=0.42]{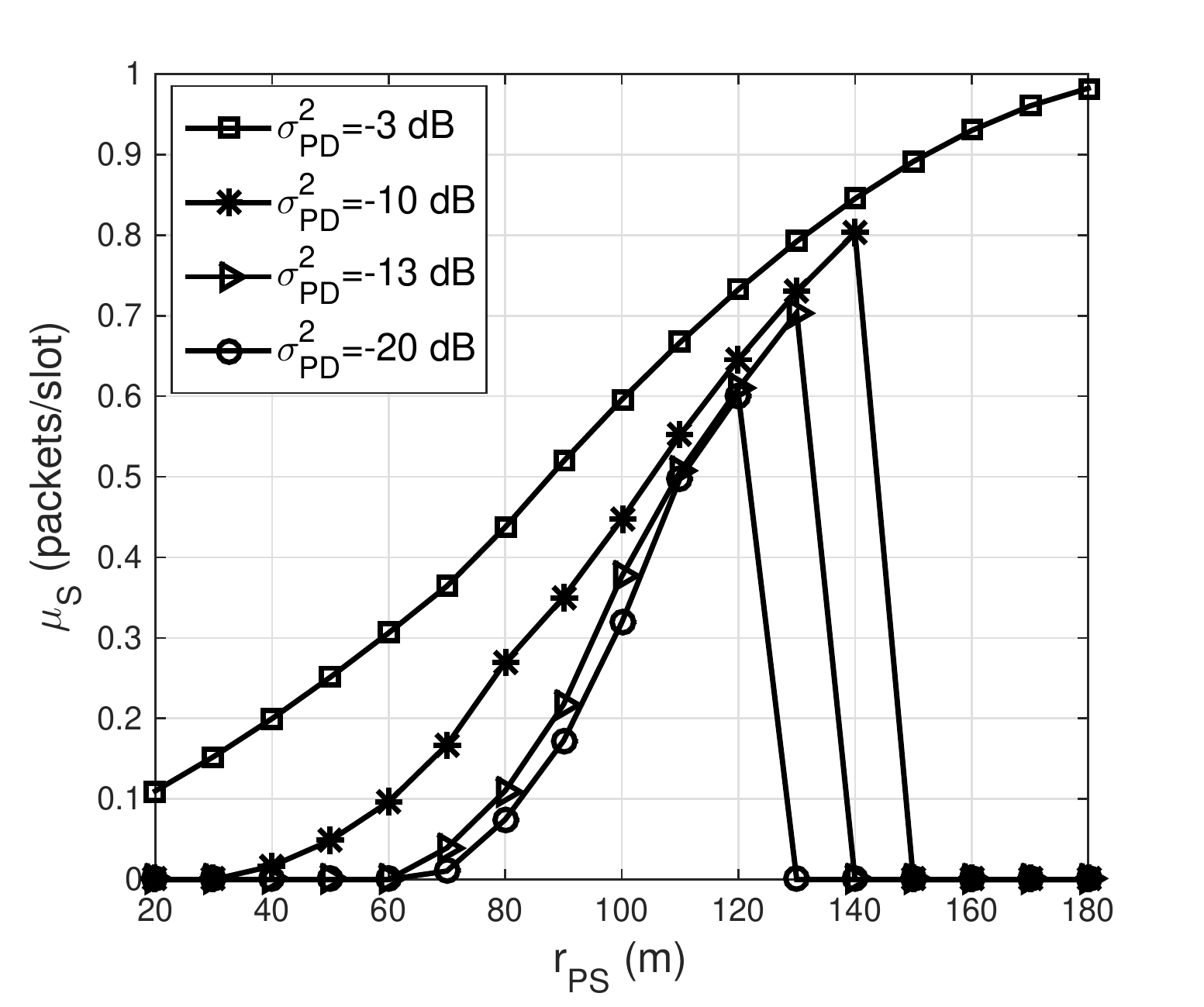} \vspace{-0.2cm}
 \protect\protect\protect\caption{Effect of distance between PU source and relay $r_{\mathcal{PS}}$
on SU packet throughput $\mu_{S}$ for $N_{P}=100$, $N_{S}=10$,
$\lambda_{P}=0.5$ and $r_{\mathcal{PD}}=200\,\mbox{m}$.\label{fig:SUvsDist}}

\vspace{-0.2cm}
 
\end{figure}

Fig. \ref{fig:SUvsDist} plots SU throughput against distance between
PU source and SU source $r_{\mathcal{PS}}$. Here, we assume that
$\mathcal{D}$ and $\mathcal{R}$ are in close vicinity and lie on
the line connecting $\mathcal{P}$ and $\mathcal{S}$. Then for given
$r_{\mathcal{PD}}$, we have $r_{\mathcal{SD}}=r_{\mathcal{SR}}=r_{\mathcal{PD}}-r_{\mathcal{PS}}$.
When $\mathcal{P}-\mathcal{D}$ channel is weak, PU packet arrival
rate at SU is high. Thus, probability of $Q_{\mathcal{S}}$ being
full is high. As SU moves away from PU source, $\theta_{\mathcal{PS}}$
decreases, while $\theta_{\mathcal{SD}}$, $\theta_{\mathcal{SR}}$,
$\overline{\theta_{\mathcal{SD}}}$ and $\overline{\theta_{\mathcal{SR}}}$
increase. This increases SU throughput. As $r_{\mathcal{PS}}$ increases
further, $\mu_{P}$ decreases to such a point that queue blocking
constraint cannot be satisfied for given $\lambda_{P}$. In this infeasible
region, $\mu_{S}$ is zero. When $\mathcal{P}-\mathcal{D}$ channel
is strong, decrease in $\mu_{P}$ due to increasing $r_{\mathcal{PS}}$
is insignificant. Thus, SU throughput $\mu_{S}$ keeps increasing
as $\mathcal{S}$ moves closer to $\mathcal{R}$.

\subsubsection{Effect of time sharing}

\begin{figure}
\centering

\includegraphics[scale=0.48]{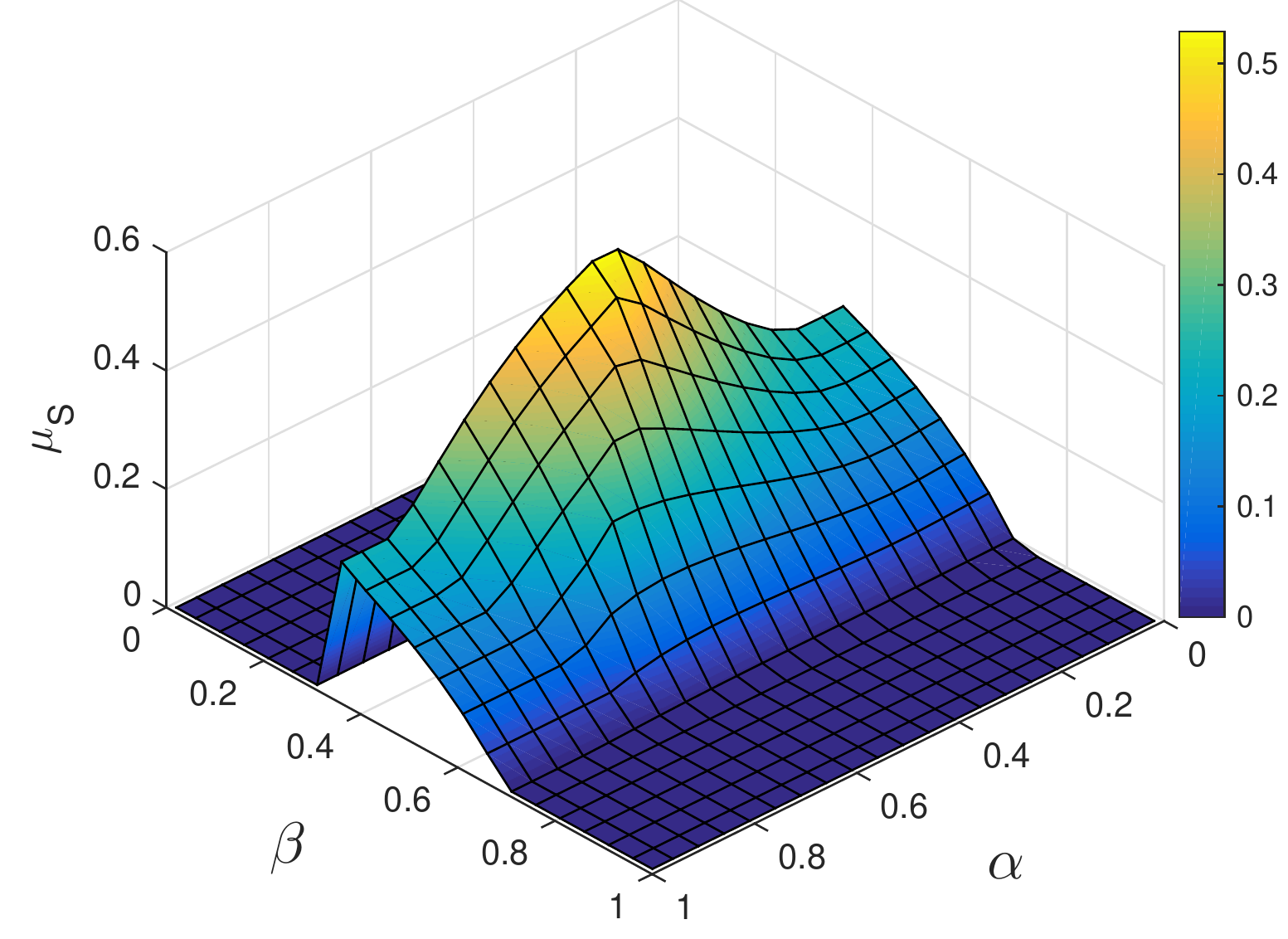} \vspace{-0.2cm}
 \protect\protect\protect\caption{Effect of time sharing factors $\beta$ and $\alpha$ on SU packet
throughput $\mu_{S}$ for $\lambda_{P}=0.5$, $N_{S}=10$, $N_{P}=50$
and $\sigma_{\mathcal{PD}}^{2}=-20\,\mbox{dB}$.\label{fig:SUvsLem_alph}}

\vspace{-0.4cm}
 
\end{figure}

Fig. \ref{fig:SUvsLem_alph} plots $\mu_{S}$ against time sharing
factors $\beta$ and $\alpha$. When $\beta$ is low, values of $\theta_{\mathcal{PD}}$
and $\theta_{\mathcal{PS}}$ are low. This results in lower value
of $\mu_{P}$ that cannot support given $\lambda_{P}$. As $\beta$
increases, PU packet departure rate increases. This allows SU to transmit
its own packets with non-zero probability. Thus, $\mu_{S}$ increases.
As $\beta$ increases further, less time is available for SU to transmit
its own packets which decreases $\theta_{\mathcal{SR}}$. Thus, $\mu_{S}$
decreases at high value of $\beta$. As $\alpha$ increases, $\overline{\theta_{\mathcal{SD}}}$
increases while $\overline{\theta_{\mathcal{SR}}}$ decreases. Increase
in departure rate of PU packets from $Q_{\mathcal{S}}$ allows SU
to transmit its own packet with higher access probability. Thus, $\mu_{S}$
increases with increasing $\alpha$. But as $\alpha$ increases further,
decrease in $\overline{\theta_{\mathcal{SR}}}$ becomes dominant,
in turn decreasing $\mu_{S}$. This indicates that there is scope
to improve $\mu_{S}$ by optimizing $\beta$ and $\alpha$. However,
the problem is difficult to solve as objective in (\ref{eq:Opt_problem})
is non-convex in $\beta$ and $\alpha$.

\section{Conclusion}

%\vspace{-0.1cm}

We studied a CR system where SU employs two-phase relaying to relay
failed PU packets. Both users have packet queues of finite capacity
which results in packet loss. We proposed optimal as well as suboptimal
access methods for SU to maximize its packet throughput under packet
loss constraint of PU. We observed that two-phase relaying model performs
better than cooperation model without time sharing for mid-range values
of PU packet arrival rate. Suboptimal methods have negligible loss
in the performance and are good low complexity alternatives to the
optimal method. Furthermore, results revealed that as relay queue
size increases, SU throughput improves initially but then decreases.
PU queue size limits maximum supported PU packet arrival rate.

%\vspace{-0.2cm}

 \bibliographystyle{IEEEtran}
\bibliography{database_full}

%\begin{IEEEbiography}{Kedar Kulkarni}%Biography text here.%\end{IEEEbiography}

%\begin{IEEEbiography}{Adrish Banerjee}%Biography text here.%\end{IEEEbiography}
\end{document}